\documentclass[12pt]{article}
\usepackage {longtable}
\usepackage {graphicx}
\usepackage[english]{babel}
\usepackage {latexsym}
\usepackage[LCY]{fontenc}
\usepackage[cp1251]{inputenc}
\begin{document}

\baselineskip=19pt

\title{Analysis of Saturn main rings by continuous
wavelet transform with the complex Morlet wavelet}

\author{E.B. Postnikov*, A. Loskutov**}

\date{\small *Theoretical Physics Department,
Kursk State University, Kursk, Radishcheva st., 33, 305000, Russia
\\ E-mail: postnicov@mail.ru \\
{**}Physics Faculty, Moscow State University, Moscow 119992,
Russia \\ E-mail: loskutov@chaos.phys.msu.ru}

\maketitle

\baselineskip=18pt

23 pages.

10 figures.

\newpage

Running head: ``Analysis of Saturn main rings ...''


\begin{abstract}

A new method based on continuous wavelet transform with the complex
Morlet wavelet to analyze Saturn main rings is presented. It allows
to investigate in detail the resonance zones and reveal the
coexistence of waves with stable periods and the wave trains with a
variable instant period. This method is based on the replacing the
integration of the fast--oscillation function by the solution of the
partial differential equations. It is shown that such an approach is
an effective tool for the study the radial structure of Saturn's A,
B, and C rings. All the analyzed images were obtained from the
Cassini spacecraft during 2004--2005 years.

\end{abstract}

\newpage

\section{Introduction}

One of the modern problems of astrophysical data processing is
development of the effective tools for the analysis of non-stationary
data sequences. As an important example of such data processing one
can mention the analysis of optical density of Saturn rings. These
possess a rich variety of patterns, waves, wakes etc. discovered by
the spacecrafts ``Voyager--1'' and ``Voyager--2'', which nature is
not completely understood up to now  (see e.g. a review of Esposito
2002). Recently,  new data become available due to the ``Cassini''
mission (see the primary report by the Porco et al. 2005). In order
to use these data and to verify various models of pattern formation,
one needs a tool to resolve local spectra.

There exists two main approaches which, can presumably, describe the
formation of such structures. The first one considers the
gravitational interaction of the ring particle with Saturn's
satellites (the basis of this approach is established by Shu et al.
1983, 1985). Within this approach, the appearance of a set of wave
trains has been explained. Later, Spilker et al. 2004 detected $40$
resonance patterns in the ring A by applying the windowed Fourier
transform. The search for the effects of high resonances in the ring
B existing on a background of stochastic perturbations, is proposed
by Thiessenhusen et al., 1995.

The second approach is based on the hydrodynamic description o the
rings (see, e.g., Schmidt and Tscharnuter 2001, Griv and Gedalin 2003
and references therein). The corresponding theoretical analysis (see
Schmidt and Tscharnuter 2001) with the use of the Navier--Stokes
equations for the viscous self-gravitating fluid reveals the radial
structures as narrow peaks divided by interstice of about 80--100
meters. Horn and Cuzzi 1996 described such a type of stable
periodicity on the basis of the ``Voyager'' data.

In the paper  Griv and Tscharnuter 2003, it has been shown that Jeans
type instability in a self--gravitating continuum may be described by
analogy with the plasma instability, where Coulomb are replaced by
the unscreened gravitational interactions. It is found that
characteristic wavelengths are about 30--200 m in the A ring, 7--30 m
in the B ring and less than 7 m in the C ring.

Thus, the local spectral properties could play a crucial role in
waves classification and in analysis of the physical properties of
the rings. The continuous complex wavelet--transform is the most
powerful tool to achieve this goal. Its principal advantage with
respect to other methods (see, e.g. Mallat 1999, Poularicas 2000) is
a high localization of the basis functions in both spatial and
frequency domain. Also, the size of a window and an instant period
are correlated in this method: for high--frequency signals the window
shrunks while for the low-frequency signals it is dilated. This keeps
the effective number of oscillations in the window constant.

The complex wavelet analysis has been successfully applied to study
the orbital period variations of asteroids in nearly--resonance zones
(Michtchenko and Nesvorn\'y, 1996) and for processing of the
solutions obtained for Hamiltonian systems, in particular for the
three--body problem (see Vela-Arevalo 2002).

An application of the wavelet transform to analize the Saturn rings
has been proposed by Bendjoya at al. 1993, who studied  the
``Voyager--2''data for the Encke gap. The authors however used only
real wavelets since their primary goal was to extract
different--scale patterns from noisy images. At the same time, the
problem of the local periodicity in the Saturn's rings requires
applications of the complex wavelet transform. The relevance of such
approach has been shown by Postnikov and Loskutov, 2005a, 2005b and
Porco et al. 2005.

In the present study we develop a new approach for the evaluation of
the complex continuous transform with the Morlet wavelet and show
that it may be successively applied for the analysis of the images of
Saturn's rings structures recently obtained during the ``Cassini''
mission. It allows to investigate in detail the resonance zones and
reveal the coexistence of waves with stable periods and the wave
trains with a variable instant period. It has a number of merits over
the standard ways based on the Fast Fourier Transform (FFT) as an
intermediate step. Firs of all, it is adapted more better to the
consideration of local features of the signal, because it never uses
the global transform of the whole sample. By the same reason, the
proposed method is free from errors related to the periodization of
the signal (the Gibbs phenomenon, ringing, aliasing, etc.). Also, it
allows to choose the boundary conditions adapted to the local
properties of functions in the region of the end points of the finite
sample. The possibility of a sufficiently small scale discretization
which satisfies a definition of the wavelet transform, allows us to
analyze the following characteristics of the Saturn main rings: to
demonstrate the coexistence of waves with stable periods and the wave
trains with a variable instant period; to raise a question about the
presence of several orders of the resonance wakes; the coexistence of
the resonance waves generated by Saturn's satellites.

\section{The complex continuous wavelet transform as a method for the image processing}

The the continuous wavelet transform (CWT), according to its
definition, reads (Mallat 1999):
\begin{equation}
\label{CWT} w(a,b) = \int\limits_{ - \infty }^{ + \infty }
{f(t)\psi^* \left( {\frac{{t - b}}{a}} \right)\frac{{dt}}{a}},
\end{equation}
where the asterisk denotes the complex conjugation. For the local
spectral analysis it is convenient to use the following
normalization:
\begin{equation}
\label{norm} \int_{-\infty}^{\infty}\left| \psi \left( {\frac{{t -
b}}{a}} \right)\right|\frac{dt}{a}=C,
\end{equation}
where $C$ is a constant.

The wavelet exploited for the signal processing, is usually the
complex Morlet wavelet:
\begin{equation}
\label{Morlet}
 \psi(\xi ) =
\frac{C}{{\sqrt {2\pi } }}e^{i\omega _0 \xi } e^{ - \frac{{\xi ^2
}}{2}},
\end{equation}
where $\omega_0$ is large enough ($\omega_0 \geq \pi$) to make the
application meaningful. The corresponding wavelet-transform $w(a,b)$
plays then a role of the local spectra in the neighborhood of the
point $b$ with period $a$. Let us illustrate this by two simple
examples.

\begin{figure}
\begin{center}
\includegraphics[width=170mm]{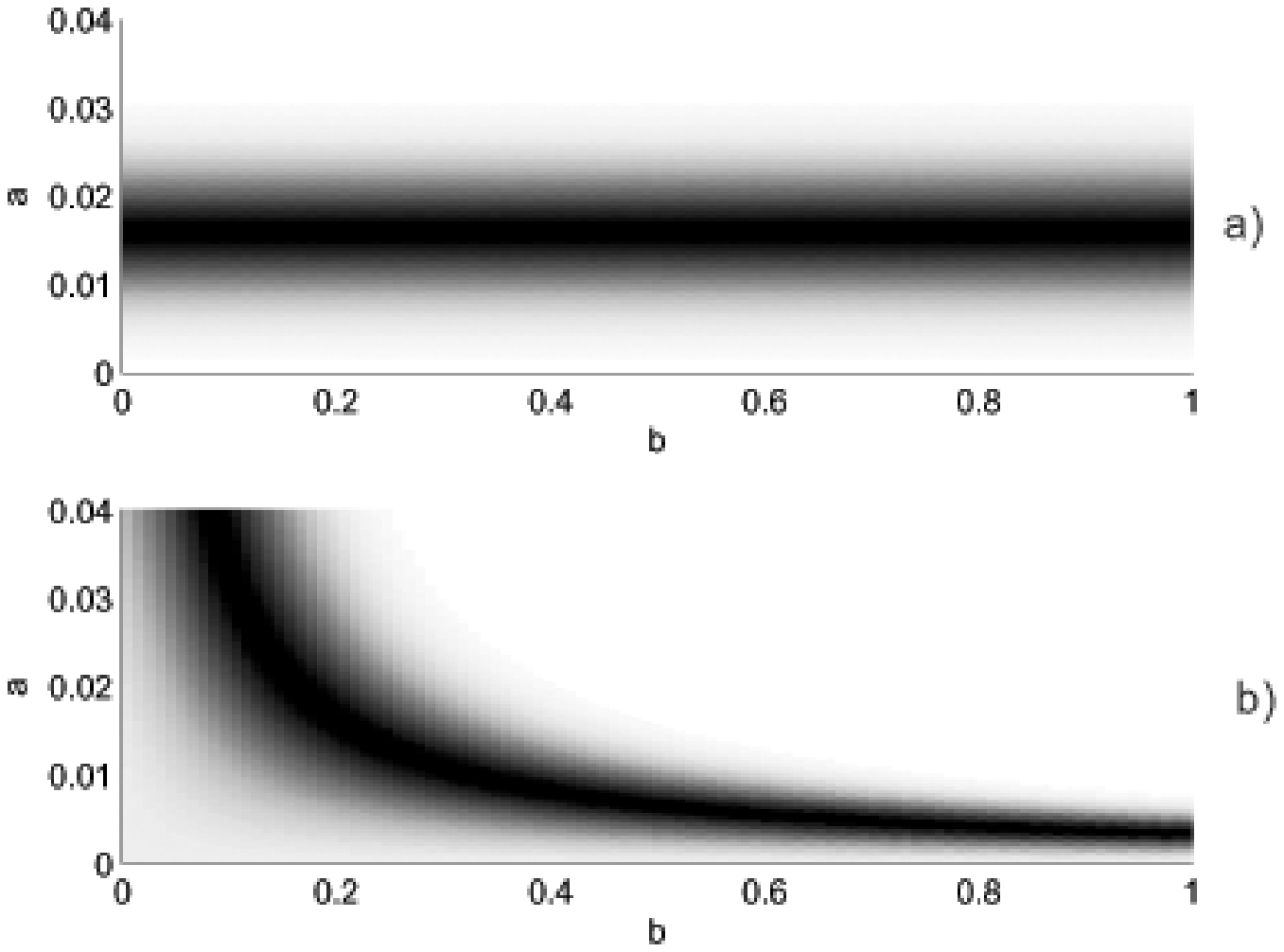}
\end{center}
\caption{The wavelet--transform of the model examples.}
\label{examples}
\end{figure}

Consider first a harmonic $\exp(\mp i\omega t)$; the corresponding
wavelet--transform and its modulus read:
$$
w(a,b) \sim e^{\pm \omega \frac{b}{a}}e^{-\frac{1}{2}(\omega_0 \pm
a\omega)^2}, \qquad |w(a,b)| \sim e^{-\frac{1}{2}(\omega_0 \pm
a\omega)^2}.
$$
The two--dimensional graph of the modulus distribution for the
wavelet--transform clearly demonstrates the maxima line corresponding
to the period $a=\pm \omega_0/\omega$, Fig.\ref{examples}a. The
Gaussian factor reduces the noise by smoothing.

One can obtain an analytic expression for the wavelet--transform
which refers to the matter distribution in the density wave due to
resonance interactions with Saturn's satellites. The asymptotic
expression in the area which are far from the perturbation source is
(Shu el al. 1983): $\displaystyle \sim\exp
\left[i\left(\frac{x^2}{2\varepsilon}-
s_\varepsilon\frac{\pi}{4}\right)\right]$. Here $s_\varepsilon=\pm 1$
and $\varepsilon$  indicates the direction of the wave motion and
$\varepsilon$ is determined by the mass distribution in the ring:
$$
\varepsilon=\frac{2\pi G \sigma}{r}\left[r\frac{d}{dr}
\left(\mu^2-(\omega-m\Omega)^2\right)\right]^{-1}.
$$
Here $G$ is a gravitational constant, $\sigma$ is the particle
surface density, $\omega$ and $\Omega$ are respectively the angular
velocity of a satellite and ring particles, $m$ is an integer, and
$\mu=\left.\frac{\partial^2 V} {\partial z^2}\right|_{z=0}$, with
$V(r,z)$ being the gravitational potential.

The modulus of the corresponding wavelet--transform reads
$$
|w(a,b)|\sim e^{-\frac{1}{4}\frac{ba\varepsilon^{-1}-
\omega_0^2}{1+a^2\varepsilon^{-1}}};
$$
it is shown in Fig.\ref{examples}b.

\begin{figure}
\begin{center}
\includegraphics[width=170mm]{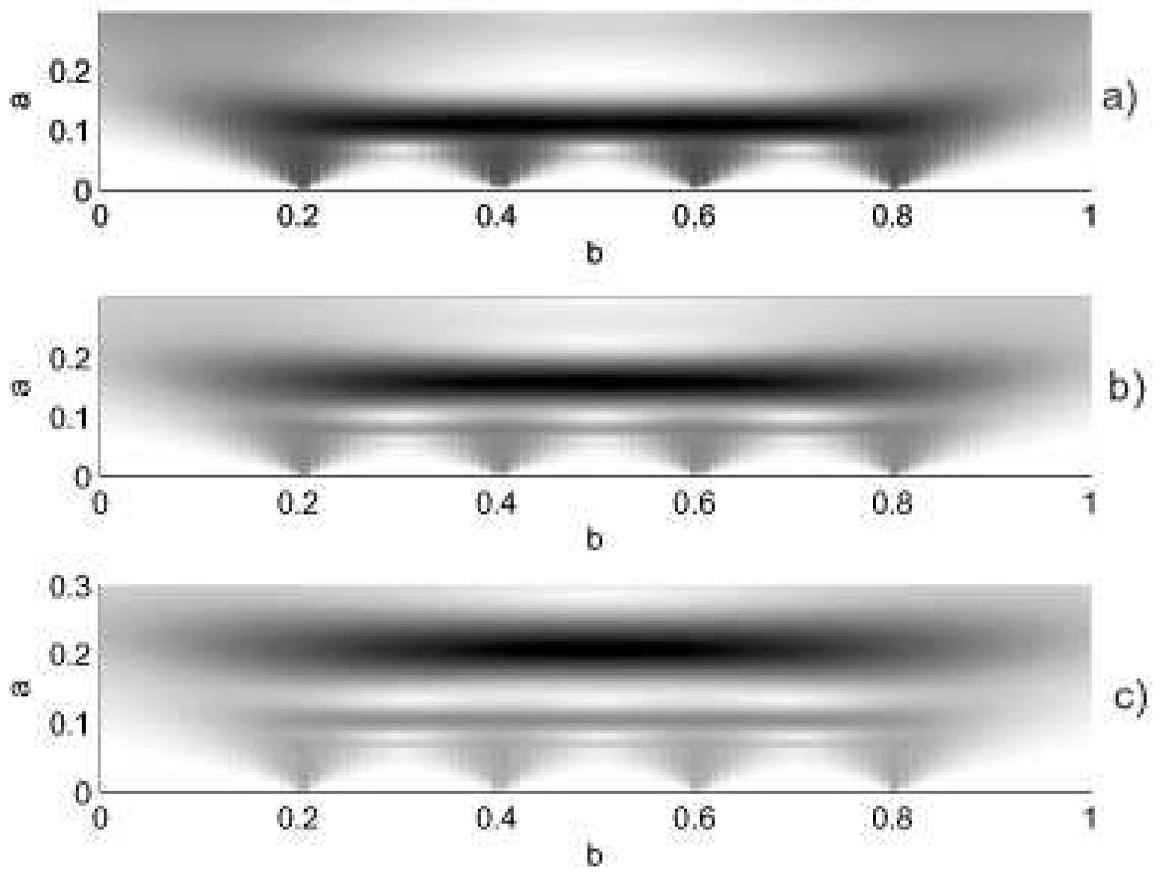}
\end{center}
\caption{Transform of the sequence of impulses in the form of the
Dirac $\delta$--functions} \label{deltawave}
\end{figure}

The variation of the basic frequency allows to vary a resolution
factor. For small $\omega_0$ the individual spikes may be easily
resolved. With increasing basic frequency, the number of oscillations
of the wavelet within a typical window also increases. However, while
the resolution of harmonics becomes better their the spatial
localization worsen. We illustrate this for a set of periodic signals
composed of Dirac $\delta$--functions:
$$
f(f)=\sum_n \delta(t-t_n),\quad w(a,b)=\sum_n \frac{C}{{a\sqrt
{2\pi } }}e^{-i\omega _0 \frac{b-t_n}{a}} e^{ - \frac{1}{2}
\left(\frac{b-t_n}{a}\right)^2},
$$
see Fig.\ref{deltawave}.

Note that we may clearly detect period of the sequence by the line
only in Fig.\ref{deltawave}a, where $\omega_0=\pi$. At
$\omega_0=1.5\pi$ (Fig.\ref{deltawave}b) one can reveal the second
line corresponding to the second harmonics in the Fourier--transform
of the signal. This phenomenon is demonstrated in
Fig.\ref{deltawave}c, where also we may see the appearance of high
harmonics.

\begin{figure}
\begin{center}
\includegraphics[width=170mm]{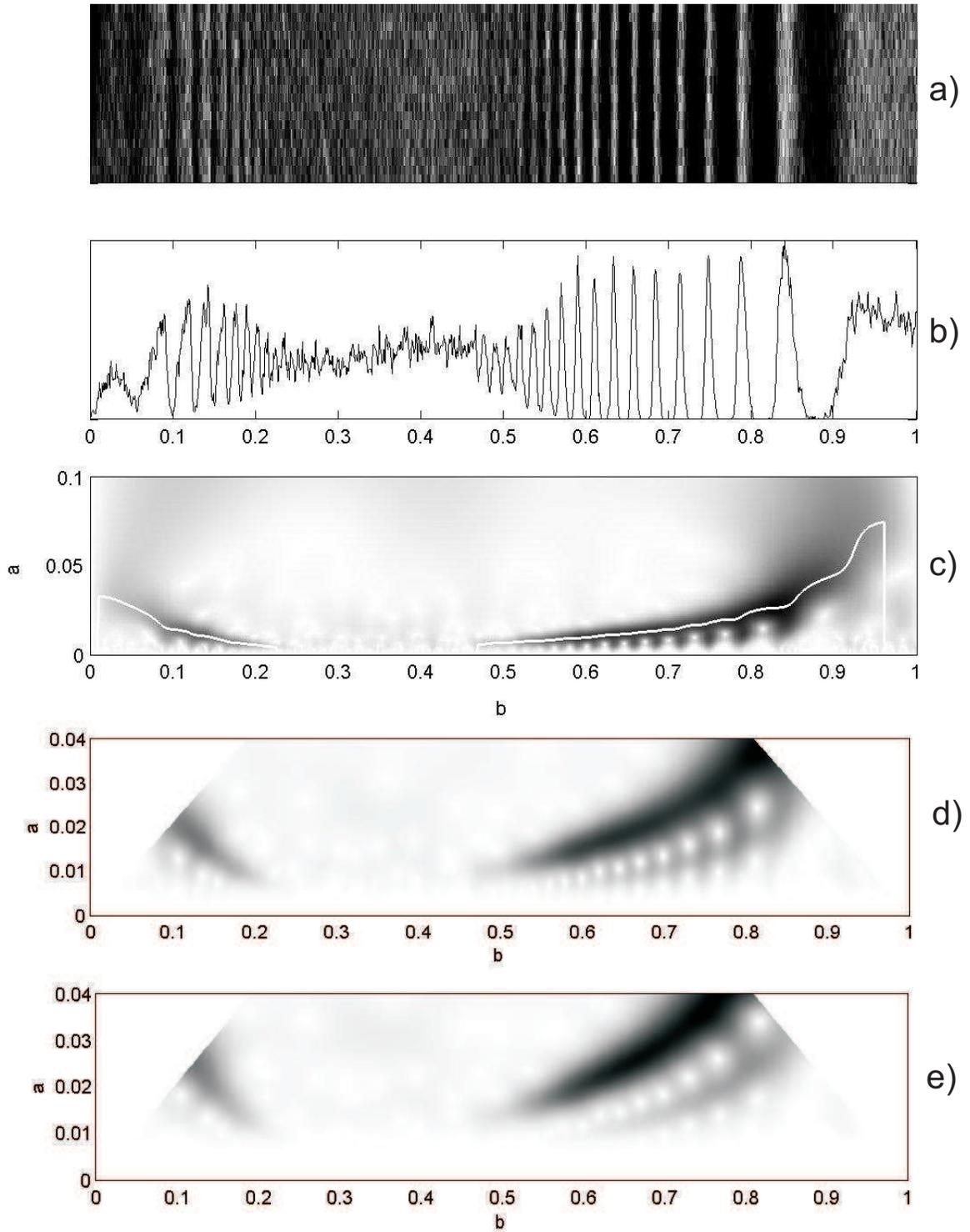}
\end{center}
\caption{The region between the resonance 12:11 with Prometheus
and 5:3 with Mimas.} \label{im4}
\end{figure}

These properties of the wavelet transform are to be taken into
account when performing the image processing. To be more specific,
consider an area of the A ring adjacent to the resonance zones due to
Prometheus (12:11) and Mimas (5:3). Owing to a large legibility of
the images the corresponding density waves may be tested to construct
the resonance excitation models (Shu et al., 1983, 1985).
Fig.\ref{im4} depicts a part of the photo obtained by ``Cassini''
(a), a contour of the image brightness (b) and the module of the
wavelet--transform at $\omega_0=\pi, \ 1.5\pi, \ 2\pi$ (c, d, e). One
can clearly see the discussed effect of the line detachment caused by
the nonharmonicity of the signal.

Such a structure has been determined by Porco et al., 2005 at the
edge of the Encke division. The authors did not perform a detailed
analysis, but assumed that the phenomenon may be explained by the the
second order perturbation due to Pan. Since the authors used a
standard calculation technique of the wavelet transform with a large
basis frequency, one needs a more careful analysis of the physical
and computational effects constituting this feature.

\begin{figure}
\begin{center}
\includegraphics[width=170mm]{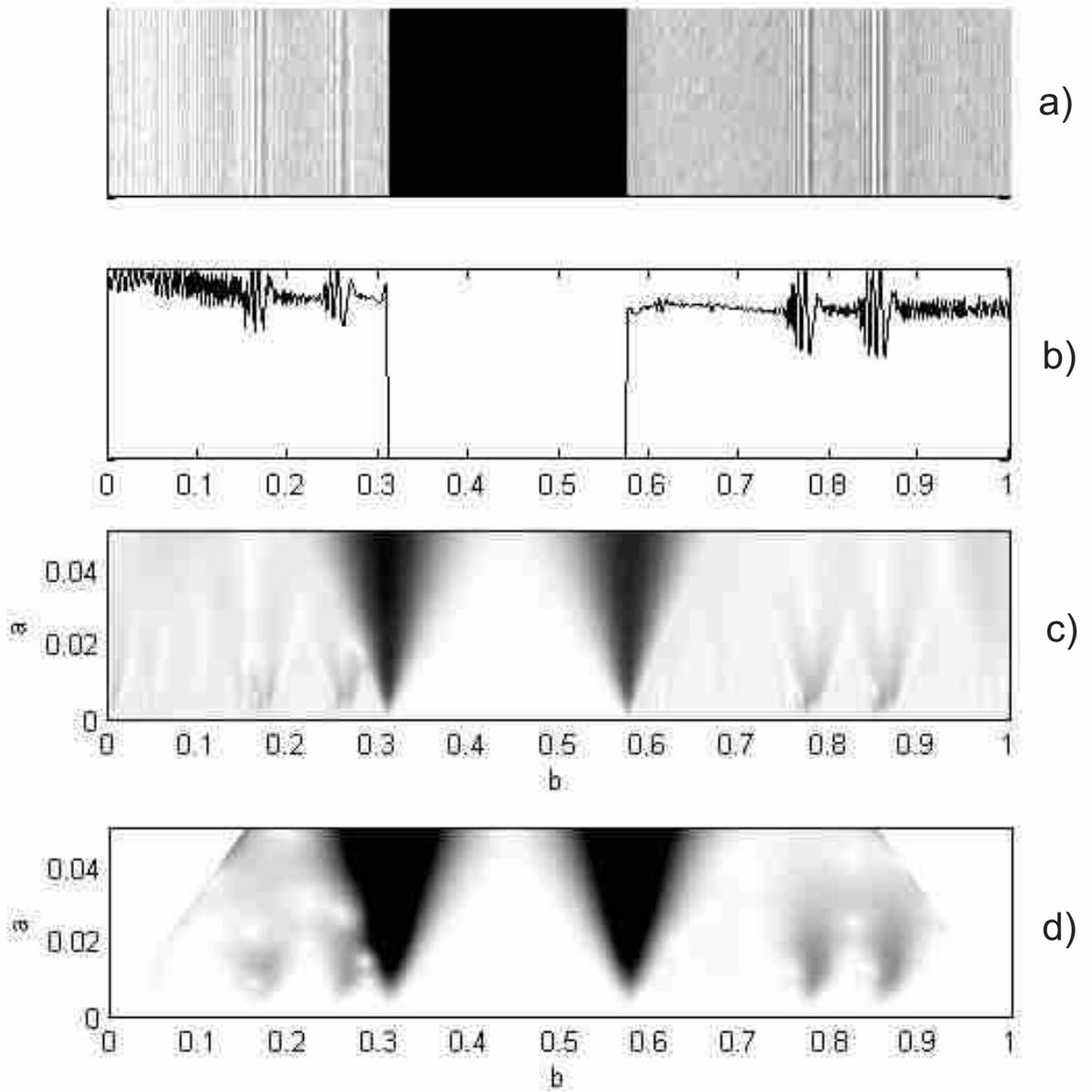}
\end{center}
\caption{The Encke division.} \label{im05-1}
\end{figure}

To process the images which lack a high resolution, one needs to
increase the basic frequency. For example, it is necessary for the
analysis of the brightness distribution of the image PIA 07533
(the Encke division, 2005, Fig.\ref{im05-1}a). Here the resolution
is about 1 km/pixel, and whole sample stretches up to about 1340
km. The brightness of the image averaged over $20$ pixels is shown
in Fig.\ref{im05-1}b. For this resolution the resonance structures
are rather narrow which generate at the basis frequency
$\omega_0=\pi$ the comb--like structures (Fig. \ref{im05-1}c).
Increasing the basis frequency up to $\omega_0=1.5\pi$ yields more
smooth maxima lines which describe typical long--wave patterns in
the interresonance region.

\begin{figure}
\begin{center}
\includegraphics[width=170mm]{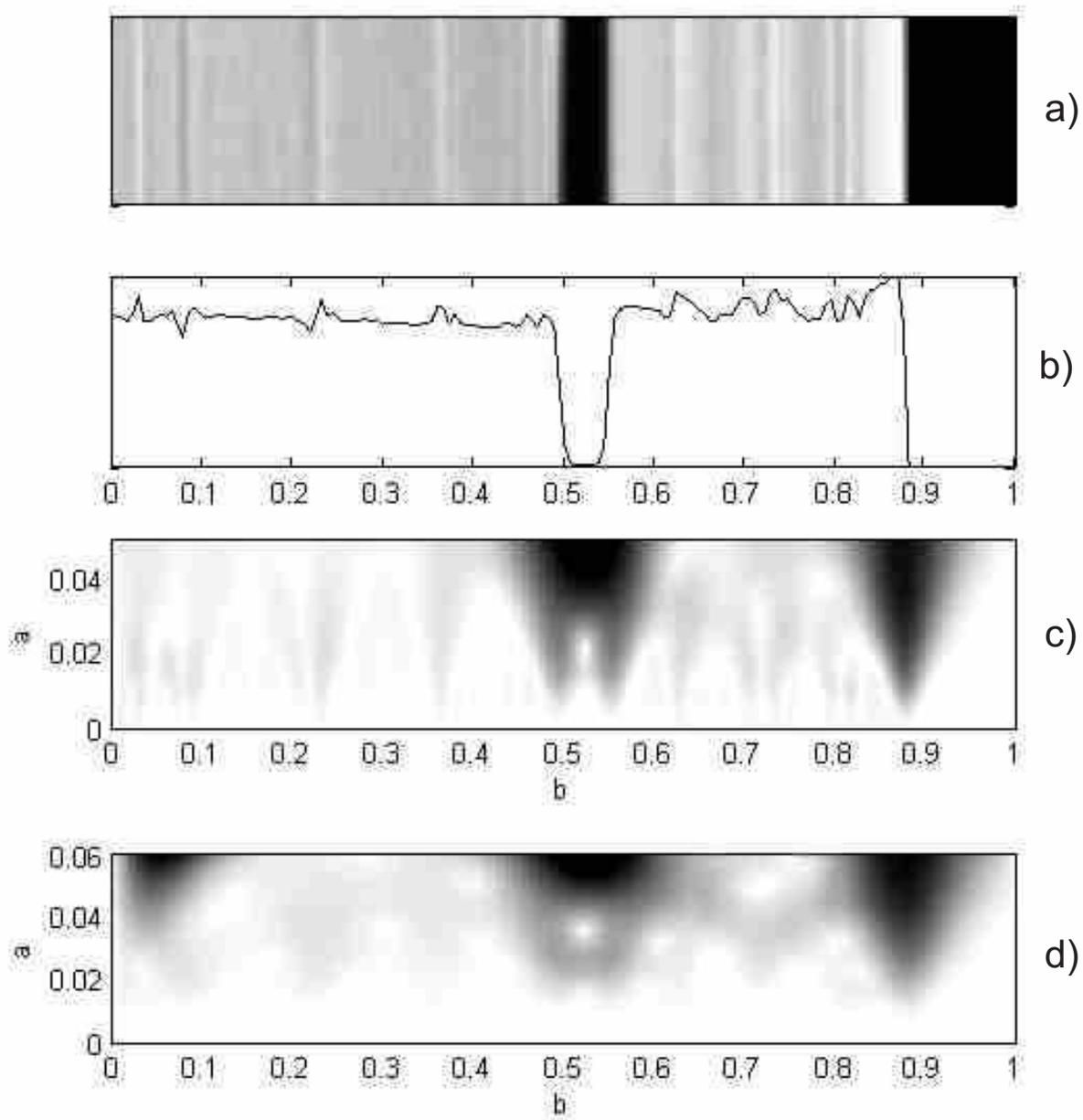}
\end{center}
\caption{The Keeler gap.} \label{im05-2}
\end{figure}

The same discussion is relevant in application to the Keeler gap
image obtained by ``Cassini'' in May 2005 (Fig.\ref{im05-2}) with the
resolution about 3 km/pixel. The size of this part is $980$ km. The
density distribution averaged  over 18 pixels in the transverse
location is shown in Fig.\ref{im05-2}b. The abrupt changes in the
image give sufficient distortions in the wavelet--transform with
frequency $\omega_0=\pi$. At the same time, the transformation with
$\omega_0=\pi$ allows to observe the evolution  of the instant period
along the radial direction.

We wish to stress that the standard way of CWT calculation includes
FFT as an intermediate step. However, in spite of the algorithm
simplicity and its hight effectiveness, the FFT suffers from certain
disadvantages: the initial sample must have $2^N$ equidistant nodes,
and the obtained data have the corresponding frequency distribution.
Violation of this condition leads to a sufficient complications of
calculation and/or to the loss of accuracy.

Here we attack the problem in a quite different way: We use the fact
that the integral (\ref{CWT}) has a form of the solution of a partial
differential equation (PDE), where the wavelet $\psi$ plays a role of
the Green function. Here the variables $a$ and $b$ correspond to the
``time'' and ``space'' variable of this PDE. Let us consider this
statement in more detail.

It is known that the wavelet--image (\ref{CWT}) with the Morlet
wavelet (see ) is a solution of the following PDE:
\begin{equation}
\label{PDE}
    \left( {a\frac{{\partial ^2 }}{{\partial b^2 }} - \frac{\partial
}{{\partial a}} - i\omega _0 \frac{\partial }{{\partial b}}}
\right)w(a,b) = 0.
\end{equation}
In \cite{Haase} this differential equation has been used to reveal
local properties of a priori known wavelet--image. As a next step we
recast the continuous transform (\ref{CWT}) with the kernel
(\ref{Morlet}) as:
\begin{equation}
\label{integral} w(a,b) = \int\limits_{ - \infty }^{ + \infty }
{f(t)\frac{{e^{ - {\textstyle{1 \over 2}}\left( {{\textstyle{{t -
b} \over a}} - i\omega _0 } \right)^2 } }}{{\sqrt {2\pi a^2 }
}}dt}.
\end{equation}
This expression corresponds to the normalization (\ref{norm}) with
$C=\exp(-\omega _0^2/2)$.

As is known, the integral (\ref{integral}) does not depends on the
imaginary subtrahend in the exponent. Also, the transformation kernel
turns to the Dirac delta-function in the limit $a \to 0$. Hence,
$w(0,a)=f(t)$ is an initial value for the PDE (\ref{PDE}). Since
$\psi(a,b)$ is a solution of (\ref{PDE}) with the initial value
$f(t)=\delta(t)$ it affirms that wavelet in (\ref{integral}) is
indeed a Green function for Eq. (\ref{PDE}).

In practical calculations it is convenient to write the
wavelet-image as a sum of real and the imaginary parts
$$
w(a,b)=u(a,b) + iv(a,b).
$$
With these notations Eqs.(\ref{PDE}) reads,
\begin{eqnarray}
\label{systemU}
\frac{\partial u}{\partial a} = a\frac{\partial ^2
u}{\partial b^2} +
\omega _0 \frac{\partial v}{\partial b}\ ,\\
\label{systemV} \frac{\partial v}{\partial a} = a\frac{\partial ^2
v}{\partial b^2} - \omega _0 \frac{\partial u}{\partial b}\ .
\end{eqnarray}
with the initial conditions
$$
\begin{array}{l}
 u(0,b) = Re(f(b)), \\
 v(0,b) = Im(f(b)). \\
 \end{array}
$$
Correspondingly, the modulus of the transform is
$$
|w(a,b)|=\sqrt{u^2(a,b)+v^2(a,b)}.
$$

From the applied point of view the proposed approach has the
following advantages.
\begin{itemize}

\item
There exist the stable finite--difference algorithms for the
numerical solution of partial differential equations of the diffusion
type (see, e.q. Sceel, Berzin 1990). The solver of parabolic PDE
realized in MATLAB (used in our examples) is based on this
algorithms. In every point $(a, b)$ for any $a$ the value of the
wavelet--transform $w(a,b)$ is determined by the solution of the
equations (\ref{systemU}). The wavelet--image obeys this system. Any
given in advance the scale small step may be easily realized by a
finite--difference scheme.

Such an approach has advantages over the standard one. The matter is
that, the method using FFT \cite{Mallat} allows to find the required
transform only for the discrete set of scales. In the other points
the value of the wavelet--transform can be calculated by the
interpolation (say, parabolic). This means that it is not satisfied
to the exact transform expression.

It is especially evident for the cases of small scales. The minimal
period resolution which is permitted by the sample of $N$ points is
$N^{-1}$. At the same time, the value of the transform at $a=0$ is
equal to zero in all points of the interval using the quadratic norm.
This norm is required for the FFT. Thus, values in the small scale
areas via the interpolation do not reveal the local properties of the
signal. In addition, the passage from the obtained result to the
amplitude norm is reduced to the division by $\sqrt{a}$ and
obviously, does not decrease the errors of the method. This is due to
the fact that this operation is ill--posed, i.e. is uses division of
two close to zero numbers. On the contrary, in the proposed approach
expression $f(b)=w(0,b)$ is an initial value for the PDE. The
solution in the small scale area possesses maximal possible accuracy.
Besides, the given method does not require equispaced sampling of the
initial values. For more details and examples, see Postnikov 2006.

\item The finite--difference algorithms do not have difficulties with the sample
periodization (the Gibbs phenomenon, ringing, and aliasing). In the
case of inequality of the function at the beginning and the end
points of the interval, one should consider the function with the
finite discontinuities in the countable set of points. For such
functions pointwise convergence of the Fourier series (the Gibbs
phenomenon) does not hold true \cite{Jerri}. As a result, in the
vicinity of the boundary points artificial oscillations (as addition
maxima in the area of small scales) appear. This is known as the
ringing phenomenon. Secondly, in the cone of influence of the
boundary points the value of errors is a quite large. This is similar
to the cones in the points of delta-functions (see Fig.
\ref{deltawave}). Depending on the value of discontinuity their
magnitude may be sufficiently more than the general level of the
modulus of the wavelet--transform of the signal. In this case
valuable information will be hidden.

At the signal periodization there is also folding of high frequency
components over a low frequency interval which is known as aliasing
\cite{Mallat}. At the inverse Fourier transform  this leads to
additional low-frequency filtering of the wavelet image. At the same
time, the algorithm basing on the PDE solution by finite--difference
schemes works with local values of the function in the knots of the
sample). For example, the standard finite--difference approximation
of the second derivative requires the knowledge of the function only
in the three points. Thus, the global characteristics of the sample
(its length and the length of the Furier--image support) does not
influence on the local properties (except for large scales compared
with half length of the sample).

Besides, owing to the reduction of the problem to the PDE system,
periodic boundary conditions are not unique. Moreover, as it was
mentioned, these conditions are undesirable. In our approach the
Dirichlet or Neuman boundary conditions are more adequate. Conditions
at every of the two ends of the interval may be chosen independently
and adapted to the properties of the signal in the vicinity of this
points.

\end{itemize}

\section{Analysis of the Saturn main rings}

\begin{figure}
\begin{center}
\includegraphics[width=170mm]{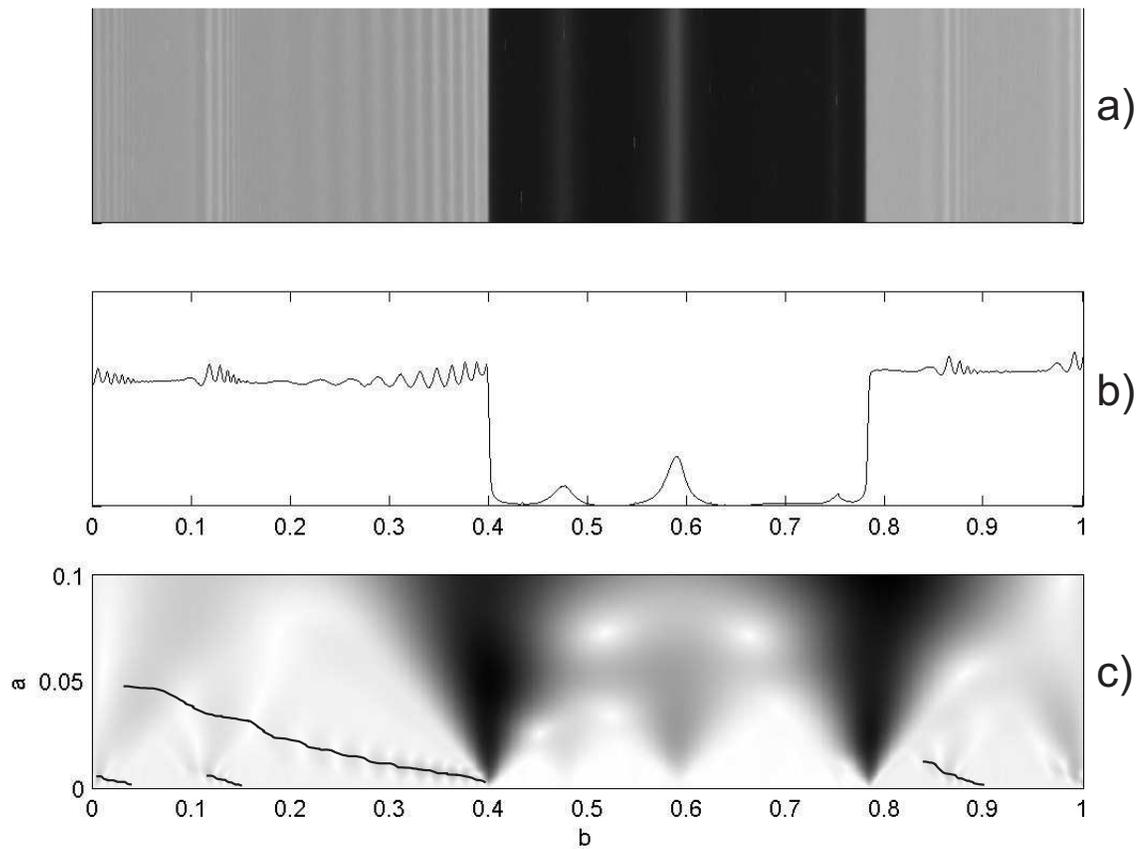}
\end{center}
\caption{The Encke division. The left side of this picture is almost
coincided with the position of the resonance 11:10 with Pandora. The
next wave--like structure is generated by the resonance 15:14 with
Prometheus. The first wave train following after this division is
generated by the resonance 12:11 with Pandora.} \label{im1}
\end{figure}

\begin{figure}
\begin{center}
\includegraphics[width=170mm]{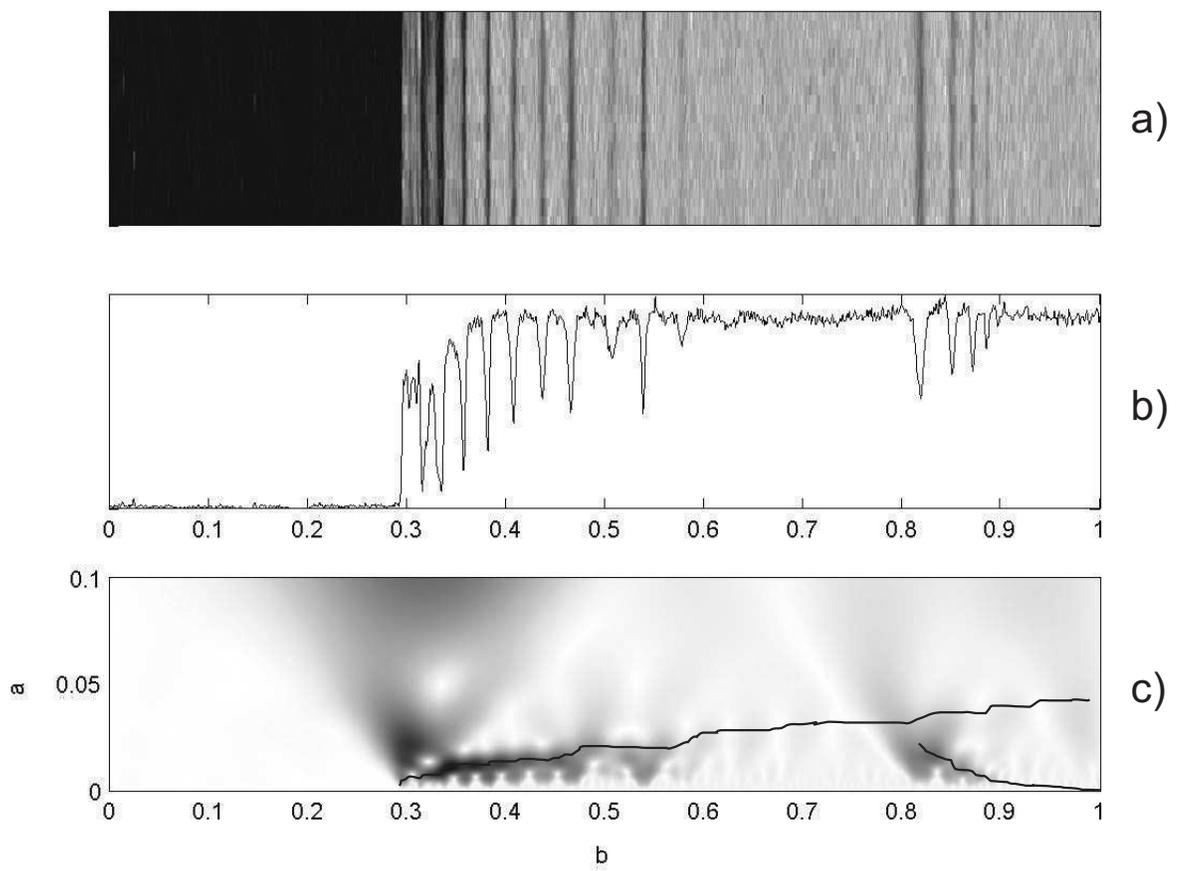}
\end{center}
\caption{The edge of the Encke division far from Saturn.} \label{im2}
\end{figure}

\begin{figure}
\begin{center}
\includegraphics[width=170mm]{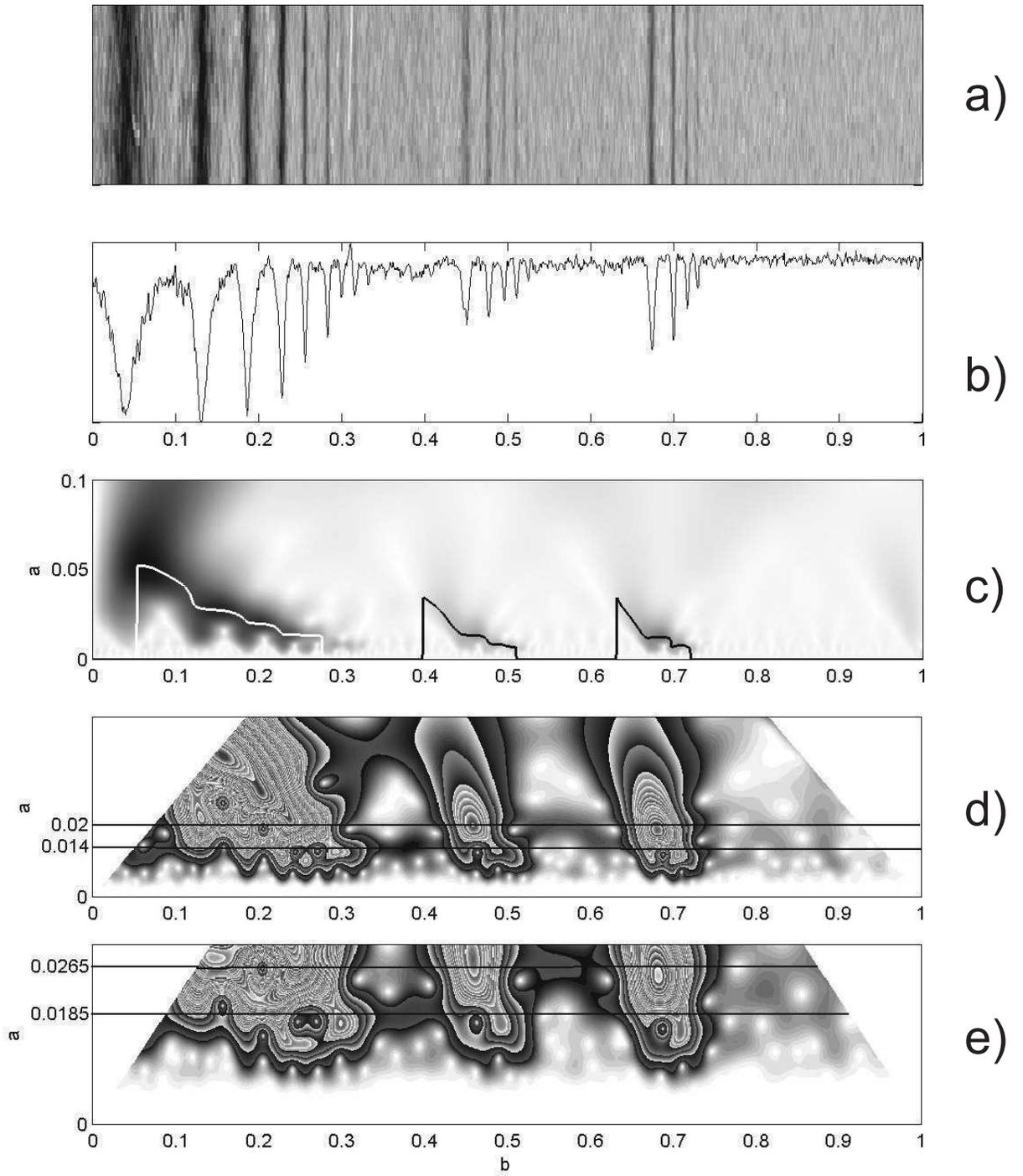}
\end{center}
\caption{Outside of A ring containing the resonance 4:3 with
Janus, 6:5 with Pandora and 7:6 with Prometheus.} \label{im3}
\end{figure}

\begin{figure}
\begin{center}
\includegraphics[width=170mm]{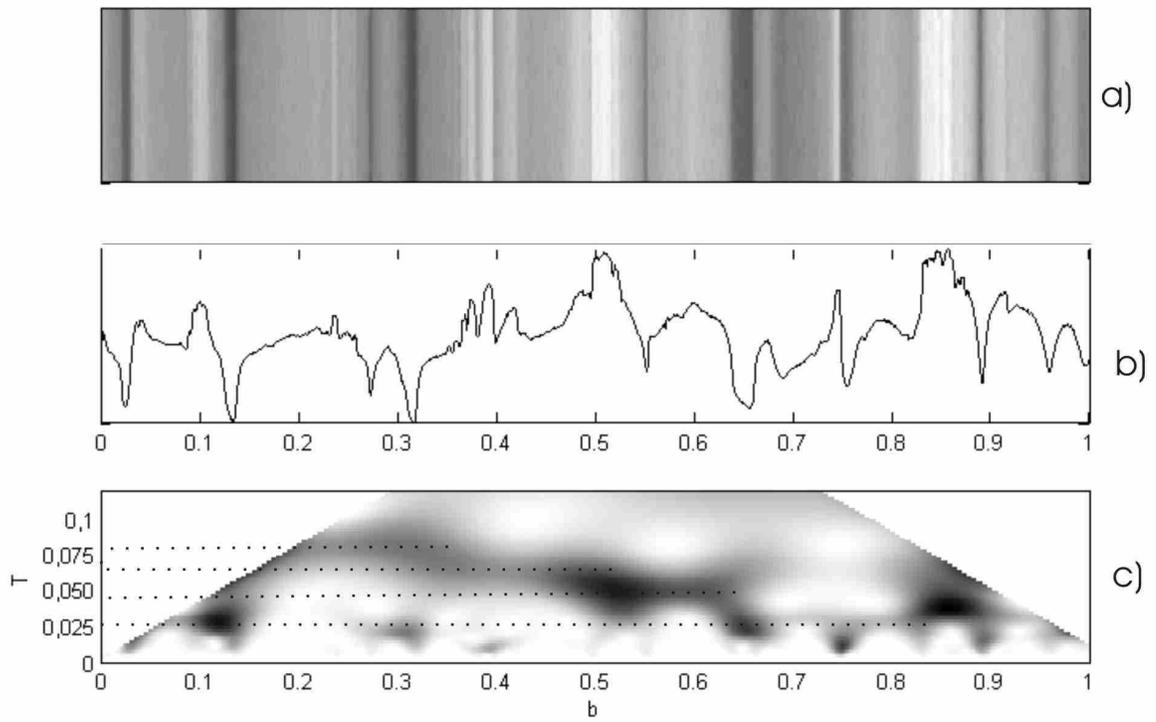}
\end{center}
\caption{The central part of B ring.} \label{ringB1}
\end{figure}

\begin{figure}
\begin{center}
\includegraphics[width=170mm]{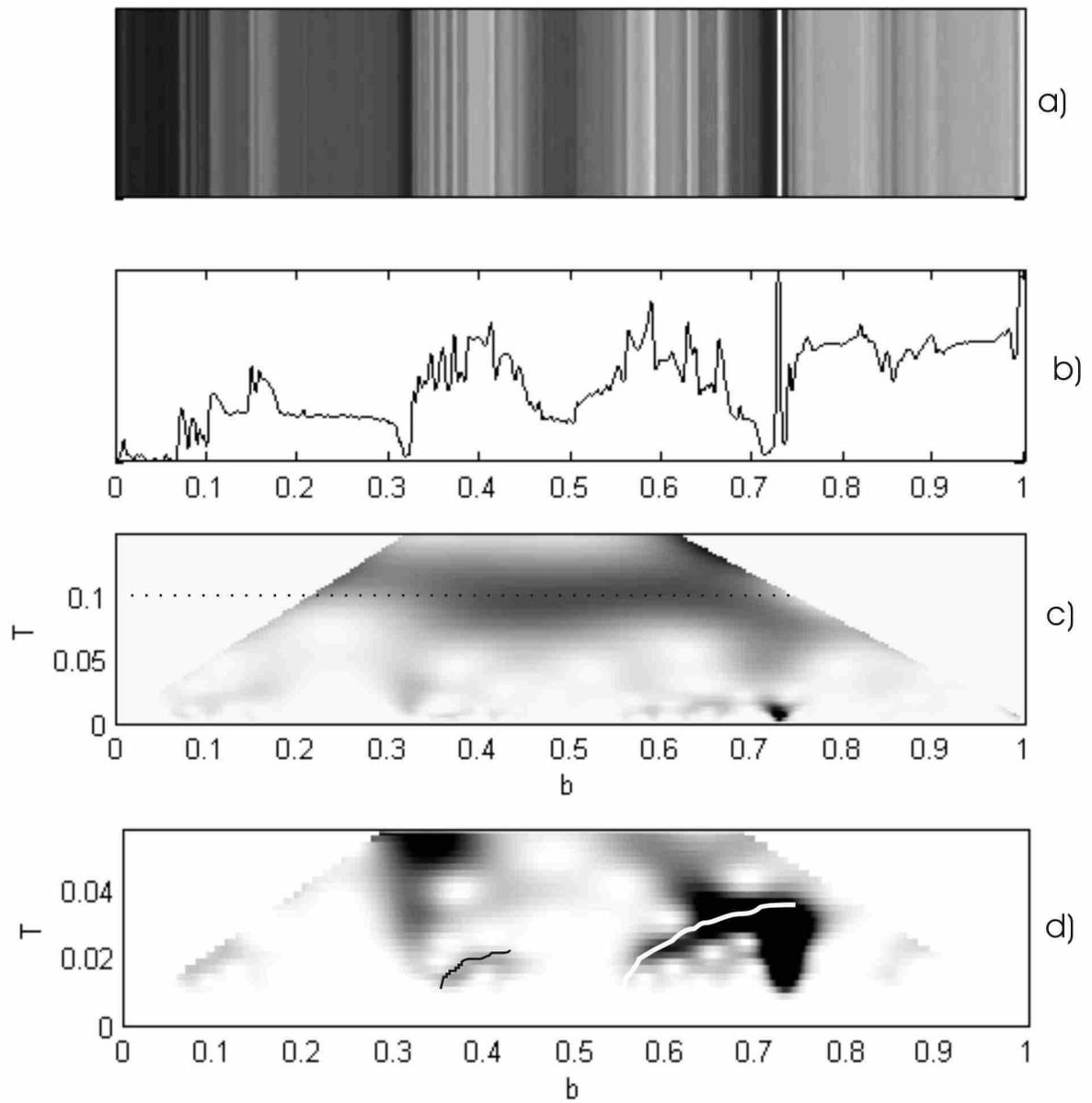}
\end{center}
\caption{A part of C ring. The distance between the center of this image and
Saturn is about 75000 km.} \label{ringC1}
\end{figure}

Let us apply the described algorithm to the processing images of
the Saturn rings obtained by the spacecraft ``Cassini''.
 We chosen images from the
NASA/JPL/Space Science Institute collection (see http://ciclops.org).
We have cut out a quite narrow stripe in the radial (across the
rings) direction from every image such that we can neglect a certain
curvature of the rings. We used the following images of A ring:
PIA06099 ($1022\times 20$ pixels, Fig.\ref{im1}a), PIA06094 ($891
\times 23$ pixels,Fig.\ref{im2}a), PIA06095 ($902 \times 23$ pixels,
Fig.\ref{im3}a) and PIA06093 ($855 \times 20$ pixels,
Fig.\ref{im4}a); B ring: PIA06543 ($1024 \times 17$ pixels,
Fig.\ref{ringB1}) and C ring: PIA06537 ($1024 \times 15$ pixels,
Fig.\ref{ringC1}). For more deep visualization all images were
stretched in the lateral direction.

These images allow a quantitative analysis since the
characteristic sizes of the ring areas may be quantified.

Due to the fact that the signal is a real, we used initial conditions
$u(0,b) = f(b)$ and $v(0,b) = 0$, where the function $f(b)$ can be
obtained by the sample average (see Fig.\ref{im1}b -- \ref{ringC1}b).
Because the signal has a bounded width the Cauchy problem for the
equations (\ref{systemU})--(\ref{systemV}) is to be replaced by the
boundary problem. Accordingly, we have chosen the boundary conditions
of the first kind: the value of the transform real part is equal to
the signal values, the transform imaginary part is zero. We used the
base frequency $\omega_0=\pi$ since it allows to select clearly lines
of the instant period, as it is shown above. In the diagrams of the
module of wavelet--images the maxima lines are marked by contrast
black lines.

An interesting fact found in the neighborhood in the Encke
division, is the co-existence of resonance waves of certain
periods. One can note that a large-scale evolution of a spiral
wave excited by Pan from the edge division, which provides the
growth of the value of its instant period, admits a continuous
transformation to a maxima line corresponding to large-scale
bursts. The typical size of such scales is about the length of the
train of resonance waves generated by the resonances 11:10 with
Pandora and 15:14 with Prometheus. Here there is a clear crossing
of the maxima lines of various resonances, which does not initiate
their interaction.

We found also the effect of the crossing of the maxima line of
large- and small-scale (forming by Pan) resonance wave structures
in the outside of the Encke division (see Fig.\ref{im2}c).

The other type of the inhomogeneity which can be detected by the
proposed wavelet--analysis is the waves with relatively stable
period. Therewith, for A ring the presence of small-scale
periodicity in interresonance areas is typical. For B and C rings
we have long-periodic waves against a background of which the
resonance peaks take place.

To carry out a detailed analysis of the A ring small-scale in the
interresonance areas, consider the density waves formed by the
resonances with Janus, Pandora and Prometheus. A typical
staircase-like shape of their instant spatial period is shown in
Fig.\ref{im3}c. obtained at the basis frequency $\pi$. To get more
frequency resolution let us increase the basis frequency up to the
value of $\omega_0=1.5$ (Fig.\ref{im3}d) and $\omega_0=2\pi$ (Fig.
\ref{im3}e). We apply the following criterium: we consider only the
areas in the plane $(b,a)$ where the condition
$\exp\left(-(b-b_0)^2/2a^2\right)\geq 10^{-5}$ is satisfied, with
$b_0=0$ or $b_0=1$.

To increase the sensitivity to the small amplitude module we use
various gray tints for larger values. This method leads to the
tailing of resonance lines but allows to separate (by the brightness
lines) almost stable periodic signal in the interval $[0.35, 0.45]$
which couples the two first resonance wave trains. Its period is $4.5
\pm 1$ km. In the interval $[0.52, 0.67]$, between the second and the
third resonances, a shortwave signal is also detected. However it has
unstable spatial period which is changed in the limits $4.5 - 6.6 \pm
1$ km. Therewith, the first value coincides with the wave period
between the resonances of Janus (4:3) and Pandora (5:6).

For B ring, in Fig.\ref{ringB1}c one can see that in the region
with dimensionless coordinates form $0.2$ to $0.65$ there exists a
smooth variation of period with $380$ to $200$ km, and the maxima
line has a staircase-like shape with three plateaus. Also, one can
see another part of the maxima line corresponding to period $115$
km. With this background and with such resolution we may find only
an inclined line of resonance in a neighborhood of the point with
the coordinate $0.4$. Small-scale structures with the ordered
periodicity are practically lacking.

One can detect a large variety of structures in a part of C ring (see
Fig.\ref{ringC1}). Here, besides a background density wave with
period 460 km we may clearly recognize two classical resonance
curves. They become apparent already in Fig.\ref{ringC1}c, being more
clearly visible in Fig.\ref{ringC1}d, where the maxima lines are
emphasized by contrast black and white curves. It should be noted
that a flat grey line, which is below this white line, does not
manifest an additional phenomena. It is just the consequence of a
large basis frequency as is shown at the analysis of non-sinusoidal
waves in the first part.

\section{Results and Discussion}

We have shown in out study, that the continuous wavelet transform
with complex Morlet wavelet is an effective tool for the analysis
of spatial radial structures of Saturn rings. It makes possible to
observe the instant period evolution at various scales as well as
detect the periodic bursts of different nature. These abilities of
the method steams essentially from the proposed algorithm, which
allows  to derive a continuous small steps for the scale variable.
It is the direct consequence of the application of the algorithm
of solving partial differential equations which replaces the
integral transform. Besides such a representation has the natural
connection with the method of tracing of wavelet maxima and ridges
offered in the paper \cite{Haase}. Additionally, the method does
not suffer, as the FFT from the artificial effects connected with
the periodization of the sample.

This makes possible to find parameters which characterize the mass
distribution as well as the other physical properties of the
rings. The detailed analysis of the wave processes in Saturn rings
should include a question concerning the interaction of long-wave
parts of perturbations with small-scale wave trains generated by
the resonance effect of
 satellites. Also, this analysis should take into account mechanisms
of the formation of almost monochromatic small-scale waves in
regions between the of resonance zones.

From the modulus diagram, we reveal that almost monochromatic
spatial density oscillations take place in such regions. They may
be detected by the maxima lines which are drawn parallel to the
coordinate axis. This line of maxima virtually connects the other
lines of maxima corresponding to two resonance staircases (i.e.
points that correspond to the minimal scale). We may propose a
mechanism of such type of behavior: when the frequency of the
resonance spatial swings of the matter is comparable with the
typical frequency of viscously-unstable standing waves, this
interaction induces visco-elastic oscillations.

The opportunity of the inverse transformation for the exact form
of the Morlet basis makes it possible to cut from the radial
signal and analyze the small-scale wave features, e.g. the
resonance trails. By substituting the wavelet--image values chosen
along the maxima line (and its small neighborhood) into the
expression of an inverse wavelet--transform, we may obtain a set
of purely resonance oscillations. Then, subtracted them from a
total signal, we get a function diagram which allows us to analyze
a small-scale structure. Localization of regions of the stable
periodicity and detection of their period in a quasihydrodynamic
model provides us a data about the density and composition of the
ring matter of the required typical viscosity caused the structure
with the given wavelength. It should be noted that the complex
wavelet--transform is much better one for such an analysis than
the other window transformation. This is due to the fact that,
owing to the self-similarity property of the analyzing wavelet, in
any frequency range one and the same typical oscillation numbers
is packed on the window width. Moreover, elimination by the
described method of a high-frequency component (this means near
zero scales) allows to denoise a signal. Finally, aperiodic spots
in the wavelet--image allows to detect structures in the form of
narrow dense rings (i.e. ringlets).

{\bf Acknowledgments}

The authors would like to thank Prof. N.Brilliantov for valuable
remarks and fruitful discussions.

\newpage

\newpage

\newpage

\newpage

\newpage

\newpage

\newpage

\newpage

\newpage

\newpage

\end{document}